\documentclass[onecolumn,showpacs,showkywords,preprintnumbers,amsmath,amsfonts,mathrsfs,amssymb,preprint]{revtex4}

\usepackage[latin1]{inputenc}
\usepackage{epsfig}
\usepackage{graphicx}
\usepackage{dcolumn}
\usepackage{bm}
\usepackage{color}
\usepackage{natbib}
\newcommand {\be}{\begin{equation}}
\newcommand {\ee}{\end{equation}}
\newcommand {\ba}{\begin{eqnarray}}
\newcommand {\ea}{\end{eqnarray}}

\usepackage{amsmath}  
\usepackage{amsfonts} 
\usepackage{graphicx} 
\usepackage{nicefrac}
\usepackage{hyperref}

\begin{document}


\title{Electronic plasma diffusion with radiation reaction force and time-dependent electric field} 
\author{J. F. Garc\'ia-Camacho${^{1,3}}$ }\email{jfgarciac@ipn.mx} 
\author{O. Contreras-Vergara${^2}$}
\author{N. S\'anchez-Salas${^2}$}\email{nsanchezs@ipn.mx}
\author{G. Ares de Parga${^2}$}
\author{J. I. Jim\'enez-Aquino${^3}$} 
\affiliation{${^1}$Departamento de Matem\'aticas,
Unidad Profesional Interdisciplinaria de 
Energ\'ia y Movilidad, Instituto Polit\'ecnico Nacional, UP Zacatenco, CP 07738, CDMX, M\'exico.
}
\affiliation{${^2}$Departamento de F\'isica,
Escuela Superior de F\'isica y Matem\'aticas, Instituto Polit\'ecnico Nacional, 
 Edif. 9 UP Zacatenco, CP 07738, CDMX, M\'exico.}
\affiliation{${^3}$Departamento de F\'isica, Universidad Aut\'onoma Metropolitana-Iztapalapa,
 C.P. 09340, CDMX, M\'exico.}

\date{\today}

\begin{abstract} 

In this work the explicit solution of the electronic plasma diffusion with radiation reaction force, under the action of an exponential decay external electric field is given. The electron dynamics is described by a classical generalized Langevin equation characterized by an Ornstein-Uhlenbeck-type friction memory kernel, with an effective memory time which accounts for the effective thermal interaction between the electron and its surroundings (thermal collisions between electrons + radiation reaction force). 
The incident electric field exerts an electric force on the electron, which in turn can induce an additional damping to the braking radiation force, allowing a delay in the electron characteristic time. This fact  allows that the effective memory time be finite and positive, and as a consequence, obtaining physically admissible solutions of the stochastic Abraham-Lorentz-like equation. It is shown that the diffusion process is quasi-Markovian which includes the radiation effects.

\end{abstract}
\pacs{05.40.-a; 52.20.Fs}

\maketitle

\section{Introduction}

The Abraham-Lorentz equation \cite{Lorentz1892,Abraham1905} is basically related to the study of 
an electron dynamics with radiation reaction force. It was derived from a Newton second law and due to 
an additional term proportional to the electron's acceleration rate of change, it represents a classical non-relativistic third-order time derivative equation. The equation leads to paradoxical solutions such as, runaway solutions and violation to causality \cite{Rohrlich65}. 
The solution of such inconsistencies goes back to the works reported in the context of classical \cite{Dirac1938,DeWitt1960,Landau1975,Rohrlich2000} and quantum 
\cite{Ford1988,Ford1991,Krivitskiui1991,OConnell2012} electrodynamics. 
It was precisely in the papers \cite{Ford1988,Ford1991,OConnell2012} in which the electronic plasma diffusion was considered as a fluctuation-dissipation phenomenon, described by quantum GLE associated with an electron embedded in a heat bath. 

In a recent paper \cite{Aresdeparga2022}, it has been shown, using the classical GLE that, the effects due to the radiation reaction force can be neglected. According to the data reported in \cite{Aresdeparga2022} it is shown that, in a classical non relativistic regime, the collision time $\tau$ is greater than the electron's characteristic time $\tau _{e}$ and thus 
the effective memory time $\hat{\tau }_{e}=\tau -\tau _{e}\simeq \tau$. Here $\tau$ is of the order of magnitude of the collision time between electrons, coming from a Brownian motion-like manner, and $\tau_e=6.26\times 10^{-24}$ 
 $s$, arises due to the radiation reaction force. So that, in this classical description, the effective friction force $m(\tau-\tau_e)\dddot x$, appearing in the Stochastic Abraham-Lorentz-like Equation (SALE), must  be $m(\tau-\tau_e)\dddot x \approx m\tau\dddot x$, and therefore, the radiation reaction force $m\tau_e \dddot x$, does not have any effect on the electronic plasma diffusion. 
Due to this fact,  it could be assumed that a classical description of the problem would not be possible. However, it will be shown in the present contribution that, it is possible to achieve the goal if the following proposal is considered. 

The idea is to make the amplitude of the effective friction force finite and positive, to take into account the influence of the radiation reaction force. This can be done by considering the incidence on the system of a time-dependent external electric field of the form $E_{ext}(t)= E_0 e^{-\omega t}$, being $E_0$ its amplitude and $\omega$ its  characteristic frequency which is proposed to satisfy $\omega={1\over k \tau_e}$, and the  constant $k$ must be such that $k\gg1$ ($k\tau_e$ is the electric field time decay). The applied electric field exerts a damping electric force on each electron (with charge $-e$) given by $F_{ext}(t)=- eE_0\, e^{-\omega t}$, 
which in turn is added to the radiation reaction force, allowing a delay in the characteristic time by $k\tau_e$. The $k\gg 1$ condition allows a control on  the value of $k\tau_e$, to be of the same order of magnitude than $\tau$, in order to have a finite and positive effective memory time, defined by $\hat\tau_{k}=\tau-k\tau_e>0$. 
This approach enables the derivation of physically consistent solutions for the SALE and avoids violation of causality.
It is worth highlighting that, even with the presence of the external electric force, the fluctuation-dissipation relation of the second kind is valid, an therefore the GLE for the velocity 
reaches its equilibrium stationary state.  Our present contribution adds to the list of works reported in the literature related to the study of Brownian motion with radiation reaction force within Classical and Quantum Mechanics, and considering gravitational effects, including the recent contribution \cite{Hsiang2022,Hsiang2019,Bravo2023,Garley2005,Garley2006,Johnson2002}.

This work is organized as follow. In Sec. II the classical GLE and its associated third order time-derivative Langevin equation are studied. The solution of the dimensionless three-order time derivative Langevin equation, is explicitly given for three specific cases of a dimensionless parameter $J$, as shown in Sec. III.
In Sec. IV, the numerical simulations are compared with theory and finally, the concluding remarks are given in Sec. V.     



\section{Theoretical approach}
 Consider the GLE associated with a free particle Brownian motion 
\be
m \dot v=-\int_0^t\gamma(t-t^{\prime})\,v(t^{\prime})\, dt^{\prime} + f(t) ,  \label{glea}  \ee
$\gamma(t)$ being the generalized friction memory kernel and $f(t)$ a Gaussian noise  with zero mean value, $\langle f(t)\rangle=0$, and a correlation function which satisfies the fluctuation-dissipation relation of the second kind \cite{Kubo1966}
\be \langle f(t) f(t^{\prime})\rangle= k_{_B}T \gamma(t-t^{\prime})  ,        \label{fdra} \ee
with $k_{_B}$ the Boltzmann's constant and $T$ the bath temperature. This fluctuation-dissipation relation guarantees the non-Markovian process (\ref{glea}) becomes stationary 
in the long time limit.

In a recent publication \cite{Aresdeparga2022}, the classical SALE for an electron gas with radiation reaction force, was derived by means of a classical GLE characterized by an Ornstein-Uhlenbeck-type friction memory kernel. In this case, the fluctuation-dissipation relation of the second kind satisfies    $\langle f(t) f(t^{\prime})\rangle= {\gamma_0 k_{_B}T\over \hat\tau_e} e^{-|t-t^{\prime}|/\hat\tau_e}$, being $\gamma_0$ the friction coefficient and $\hat\tau_e=\tau-\tau_e$, an effective memory time. However, it has been demonstrated that within this classical non-relativistic regime, where $\tau\gg\tau_e$, the effects induced by the radiation reaction force are imperceptible. To incorporate these effects into a classical description of the electronic plasma diffusion, we propose considering the influence of an external time-dependent electric field, the purpose of which is to attenuate the electron's characteristic time,  as explicitly discussed in the introduction of this work. The classical GLE is thus written as  
\be
m \dot v=-\int_0^t\gamma(t-t^{\prime})\,v(t^{\prime})\, dt^{\prime} -F_0\, e^{-\omega t} +f(t) ,  \label{gleb}  \ee
being $F_0=e E_0$. On the other side, due to the exponential decay of the external electric force, the fluctuation-dissipation relation of the second kind is also valid, and thus the stochastic process (\ref{gleb}) is stationary in the long time limit. The electric force must induce a delay in the electron's characteristic time by a quantity $k\tau_e$, 
which allows to get a finite memory time $\hat\tau_k=\tau-k\tau_e>0$, for an appropriate value of $k\gg 1$. As a consequence of this fact, the friction memory kernel is assumed to satisfy $\gamma(t-t^{\prime})={\gamma_0\over \hat\tau_k} \,e^{-|t-t^{\prime}|/\hat\tau_k}$,
and therefore the second kind fluctuation-dissipation relation reads
\be
 \langle f(t) f(t^{\prime})\rangle= {\gamma_0 k_{_B}T\over \hat\tau_k}  e^{-|t-t^{\prime}|/\hat\tau_k} .  \label{fdrb}   \ee
So, the classical GLE associated with an electron into an electron gas with radiation reaction force becomes  
\be
 m\dot v=-\frac{\gamma_0}{\hat\tau_k}\int_0^t e^{-(t-t^{\prime})/\hat\tau_k}\, v(t^{\prime})\, dt^{\prime} - F_0e^{-\omega t}+ f(t) . \label{glec} \ee
Using the provided expressions in \cite{Aresdeparga2022}
\ba
 {\eta}(t)&=&-\frac{\gamma_0}{\hat\tau_k}\int_0^t e^{-(t-t^{\prime})/\hat\tau_k} \, v(t^{\prime})\, dt^{\prime}+f(t), \label{eta}\\
 \nonumber\\
 f(t)&=&{\sqrt{\lambda}\over\hat\tau_k} \int_0^t e^{-{(t-t^{\prime})/\hat\tau_k}}\, \xi(t^{\prime})\, dt^{\prime} ,  \label{f} \ea
where $\lambda=\gamma_0 k_{_B}T$, and $\xi(t)$ is  a Gaussian white noise with zero mean 
and a correlation function  $\langle \xi(t) \xi(t^{\prime})\rangle=2 \delta(t-t^{\prime})$, it becomes straightforward to derive   
\be m \hat\tau_k {\dddot x} +m{\ddot x}=-\gamma_0 v + c_0 F_0 \, e^{-\omega t}+
\sqrt{\lambda}~\xi(t)  , \label{sale1} \ee
with $c_0=\omega\tau-2$. This is a classical SALE which describes the electronic plasma diffusion taking into account the braking to radiation force, when the system is under the action of an exponential decay electric field. In terms of dimensionless variables $y=c_1 x$  and $s=c_2 t$, with $c_1={m\over\hat\tau_{k}^2 c_0 F_0}$ and $c_2={1\over \hat\tau_k}$, Eq. (\ref{sale1}) is further expressed as
\be
{d^3 y\over ds^3}+{d^2 y\over ds^2} + J {d y\over ds}
={1\over c_0} e^{-(\omega\tau-1)s}+ {\gamma_0 \over c_0 F_0} \sqrt{D} ~ \xi(\hat\tau_k\, s) , \label{sale2}  \ee
being $J={\hat\tau_k\over \tau_r}$, with
$\tau_r={m\over \gamma_0}$, the relaxation time, and $D={k_{_B}T\over \gamma_0}$, Einstein's diffusion coefficient. The product $\omega\tau={\tau\over k\tau_e}>2$, accounts for the  coupling effect between both the memory time $\tau$ and scaled electron characteristic time $k\tau_e$. In the next section we will calculate the statistics of the electron gas Brownian motion by means of the solution of Eq. (\ref{sale2}).


\section{Explicit solutions}

The explicit solution of Eq. (\ref{sale2}) has three roots given by   
\be
\lambda_1=0, \qquad \lambda_2=-{1\over 2}+{1\over 2}\sqrt{1-4J}, \qquad \lambda_3=-{1\over 2}-{1\over 2}\sqrt{1-4J} , \label{l23}  \ee
from which also three cases can be analyzed, namely: the cases of real, complex, and critical roots. Therefore, this highlights the essential role played by the J parameter.   

\subsection{The case of real roots $1-4J>0$} 

This case implies that $J<{1\over 4}$, which means that $\hat\tau_k<{1\over 4}\tau_r$. Taking into account that $\hat\tau_k=\tau-k\tau_e= {1\over\omega}(\omega\tau -1)$ and 
$\omega\tau>2$, thus 
$2<\omega\tau<1+{1\over 4}\omega\tau_r$. For zero initial conditions, the solution for the average of the dimensionless quantities $y$, $v_y$, and $a_y$, being  
$v_y={d y\over ds}$, and $a_y={d v_y\over ds}$, are shown to be  
\ba
\langle y \rangle
&=&{1-e^{-(\omega\tau-1)s} \over \lambda_2\lambda_3(\omega\tau-1)}
- {e^{-\lambda_2 s}-e^{-(\omega\tau-1)s} \over {\lambda_2(\lambda_3-\lambda_2)(\lambda_2+\omega\tau-1)}} \cr\cr
&+&{e^{-\lambda_3 s}-e^{-(\omega\tau-1)s} \over {\lambda_3(\lambda_3-\lambda_2)(\lambda_3+\omega\tau-1)}}, \label{my1}  \\
\nonumber\\
\langle v_y \rangle
&=&{-{e^{-\lambda_2 s}-e^{-(\omega\tau-1)s} \over {(\lambda_3-\lambda_2)(\lambda_2+\omega\tau-1)}}
+{e^{-\lambda_3 s}-e^{-(\omega\tau-1)s} \over (\lambda_3-\lambda_2)(\lambda_3+\omega\tau-1)}}, \label{mvy1}  \\
\nonumber\\
\langle a_y \rangle
&=&{-\lambda_2(e^{-\lambda_2 s}-e^{-(\omega\tau-1)s}) \over {(\lambda_3-\lambda_2)(\lambda_2+\omega\tau-1)}}
+{\lambda_3 (e^{-\lambda_3 s}-e^{-(\omega\tau-1)s}) \over {(\lambda_3-\lambda_2)(\lambda_3+\omega\tau-1)}}. \label{may1}  \ea
Also, upon the definition of dimensionless fluctuating variables $Y=y-\langle y\rangle$, $V_y=v_y-\langle v_y\rangle$, $A_y=a_y-\langle a_y\rangle$, it can be shown that the variances 
$\sigma^2_Y=\langle Y^2\rangle$, $\sigma^2_{V_y}=\langle V^2_y\rangle$, and $\sigma^2_{A_y}=\langle A^2_y\rangle$, are shown to satisfy 
\ba
\langle Y^2 \rangle
&=&{\gamma_0^2 D\over \hat\tau_k c_0^2 F_0^2} \bigg\{ {2s \over (\lambda_2\lambda_3)^2 }+ { e^{2\lambda_2 s}-1 \over \lambda_2^3 (\lambda_3-\lambda_2)^2} +{e^{2\lambda_3 s}-1 \over \lambda^2_3 (\lambda_3-\lambda_2)^2} \cr\cr
&+&4\bigg[ {e^{\lambda_3 s}-1\over \lambda_2\lambda_3^3(\lambda_3-\lambda_2)}-{e^{\lambda_2 s}-1\over \lambda_3\lambda_2^3(\lambda_3-\lambda_2)}-{e^{(\lambda_3+\lambda_2)s}-1\over \lambda_3\lambda_2(\lambda_3-\lambda_2)^2(\lambda_3+\lambda_2)} \bigg] \bigg\}, \label{Y2}  \\ 
\nonumber\\
\langle V_y^2 \rangle
&=&{\gamma_0^2 D\over \hat\tau_k c_0^2 F_0^2} {1\over(\lambda_3-\lambda_2)^2} \bigg[{e^{2\lambda_2s}-1\over \lambda_2}-4{e^{(\lambda_3+\lambda_2)s}-1\over \lambda_3+\lambda_2}+{{e^{2\lambda_3s}-1}\over{\lambda_3}} \bigg], \label{Vy2}  \\
\nonumber\\
\langle A_y^2 \rangle
&=&{\gamma_0^2 D\over \hat\tau_k c_0^2 F_0^2} {1\over(\lambda_3-\lambda_2)^2}\bigg[\lambda_2(e^{2\lambda_2s}-1)-4{{\lambda_2\lambda_3(e^{(\lambda_3+\lambda_2)s}-1)}\over{\lambda_2+\lambda_3}}+\lambda_3(e^{2\lambda_3s}-1)\bigg]. \label{Ay2}  \ea

We can return to the original variables using again the transformations $y=c_1 x$, $s=c_2 t$, and also the fluctuating variables $X=x-\langle x\rangle$, $V=v-\langle v\rangle$, and $A=a-\langle a\rangle$. In particular, in the limit of long time the variances in the original variables become 
\be 
\langle X^2\rangle =2D\left(t+{J-3 \over 2}\tau_r\right) , \qquad 
\langle V^2\rangle_{eq}={k_{_B}T\over m},\qquad  \langle A^2\rangle_{eq}={1\over J\, \tau_r^2} \langle V^2\rangle_{eq} .
\label{msdxva} \ee
The expression $\langle V^2\rangle_{eq}={k_{_B}T\over m}$, is the equilibrium expected result, and the acceleration variance is proportional to $\langle V^2\rangle_{eq}$, and $J$-dependent.   
However, due to the fact that 
$0<J<{1\over 4}$, it can be neglected respect to number 3 and thus 
the variance $\langle X^2\rangle$ can be approximated by  
\be \langle X^2\rangle =2D\left( t- {3\over 2}\tau_r\right), \label{x22}
\ee
which is independent of the $J$ parameter. The comparison of these theoretical results with the numerical simulation of Eq. (\ref{sale2}), is given in detail in Sec. IV.

\subsection{The case of complex roots $1-4J<0$ } 

 In this case $J>{1\over 4}$ or $\hat\tau_k>{1\over 4}\tau_r$, which means that $\omega\tau >1+{1\over 4}\omega\tau_r$. The two complex roots can be defined as  $\lambda_2=a+{\rm i} b$, and $\lambda_3=a-{\rm i} b$, being $a=-{1\over2}$ and ~$b={1\over2} \sqrt{4J-1}$. So, for zero initial conditions, the mean values $\langle y\rangle$, $\langle v_y\rangle$, and $\langle a_y\rangle$, can be written as 
\ba
\langle y \rangle &=& {1\over J} \big[ I_0+ K_1 I_1 +K_2I_2\big] ,  \label{my2} \\
\langle v_{y} \rangle &=& {1\over J}\big[ {-K_1^{\prime} I_1 + K_2^{\prime} I_2} \big], \label{mvy2} \\
\langle a_{y} \rangle &=&{1\over J} \big[-K_1^{\prime\prime} I_1+ K_2^{\prime\prime}  I_{2} \big] ,  \label{may2}\ea
where $K_1^{\prime}$ and $K_2^{\prime}$ are the derivative respect to $s$ variable, and 
\ba
I_0&=&{1\over\omega \tau}\big[ 1-e^{-\omega \tau s} \big] , \label{I0} \\
I_1&=&{1\over (a+ \omega \tau)^2+b^2}
\big\{  b- e^{-s( \omega \tau +a)} \big[b \cos bs
+ (a+\omega\tau) \sin bs \big] \big\} ,  \label{I1} \\
I_2&=& {1\over (a+ \omega \tau)^2+b^2}
\big\{ a+\omega\tau + e^{-s( \omega \tau +a)}\big[b \sin bs
- (a+\omega\tau) \cos bs \big] \big\} , \label{I2}\\
K_1&=&e^{as}\big[\frac{a}{b} \cos bs + \sin bs \big], \label{K1}\\
K_2&=&e^{as}\big[\frac{a}{b} \sin bs -\cos bs\big]. \label{K2}  \ea
The variances have respectively the following expressions 
\ba 
\langle Y^2 \rangle&=&{\gamma_0^2 D\over \hat\tau_k c_0^2 F_0^2}{1\over J^3}
\big[2J s + K_1^2\, {\mathcal B} + K_2^2\, {\mathcal C} - 2\big(K_1\, {\mathcal D} - K_2\,{\mathcal E} + K_1K_2\, {\mathcal F}\big) \big], \label{Y2a} \\
\langle V_y^2\rangle &=& {\gamma_0^2 D\over \hat\tau_k c_0^2 F_0^2}{1\over J^3}
\big[K_1^{\prime\,2}\, {\mathcal B} -2 K_1^{\prime} K_2^{\prime}\, {\mathcal F} +K_2^{\prime\,2}\,{\mathcal C}\big] , \label{Vy2a}\\
\langle A_y^2\rangle&=&{\gamma_0^2 D\over \hat\tau_k c_0^2 F_0^2}{1\over J^3}\big[ K_1^{\prime\prime\,2} \,{\mathcal B} -2K_1^{\prime\prime}K_2^{\prime\prime}
\, {\mathcal F} + K_2^{\prime\prime\,2}\, {\mathcal C} \big],  \label{Ay2a}    \ea
with 
\ba
{\mathcal B}&=&-\big[b^2+e^{-2as}\big(a^2\cos 2bs-ab\sin2bs-J \big)  \big],  \label{I12} \\
{\mathcal C}&=&-\big[a^2+J-e^{-2as} \big(a^2\cos2bs-ab\sin 2bs +J\big)  \big],  \label{I22} \\
{\mathcal D}&=&\big[2b -2e^{-as} \big(a\sin bs +b\cos bs \big)\big],  \label{I0I1} \\
{\mathcal E}&=& \big[ 2a -2e^{-as}\big(a\cos bs - b\sin bs\big)  \big],  \label{I0I2} \\
{\mathcal F}&=&{1\over 2}  \big[ b- e^{-2as}\big(a\sin 2bs +b\cos 2bs\big) \big].  \label{I1I2} \ea
Also in the long time limit, the variances $\langle X^2\rangle$, $\langle V^2\rangle$ and $\langle A^2\rangle$ are the same as those given in Eq. (\ref{msdxva}). In this case, the variance $\langle X^2\rangle$ is valid for all  $J>0.25$, and in  particular for $J=3$, the variance reduces to $\langle X^2\rangle=2Dt$, which is the same as the Markovian result. The details are also given in Sec. IV.
    

\subsection{The critical case $1-4J=0$} 

It is clear in this case that $J={1\over 4}$ or $\hat\tau_k={1\over 4}\tau_r$. So, the dimensionless variables now read $y={16 m\over \tau_r^2 c_0F_0} x$, $s={4\over \tau_r} t$, and $c_0={1\over 4} (\omega\tau-4)$, and therefore, the solution of the corresponding dimensionless stochastic differential equation leads to the following  statistics 
\ba 
\langle y \rangle&=&{ 4[1-e^{-(\omega \tau-1)s}]\over \omega \tau-1}
-{2[s+2][1-e^{-(\omega\tau-{3\over2})s}] e^{-s/2}\over\omega\tau-{3\over2}} \cr\cr
&+&{2\{1+[(\omega\tau-{1\over2})s-1]e^{-(\omega \tau-{3\over2})s} \} e^{-s/2}\over(\omega \tau-{3\over2})^2}, \label{myc} \\
\nonumber\\
\langle v_y \rangle&=&{ s[1-e^{-(\omega \tau-{3\over 2})s}] e^{-s/2}\over \omega\tau-{3\over 2}} - {\{1+[(\omega\tau-{1\over2})s-1]e^{-(\omega \tau-{3\over2})s} \} e^{-s/2}\over(\omega \tau-{3\over2})^2} , \label{mvyc}\\ 
\nonumber\\
\langle a_y \rangle&=&{ (1-{s\over2})[1-e^{-(\omega \tau-{3\over2})s}] e^{-s/2}\over \omega\tau-{3\over 2}} - {\{1+[(\omega\tau-{1\over2})s-1]e^{-(\omega \tau-{3\over2})s} \} e^{-s/2}\over 2(\omega \tau-{3\over2})^2} . \label{mayc}
\ea
The variances are shown to be 
\ba 
\langle Y^2 \rangle&=&{\gamma_0^2 D\over \hat\tau_k c_0^2 F_0^2}\big[
32s-176 +64(s+4)e^{-s/2}-8(s^2+6s+10)e^{-s} \big] , \label{Y2c}\\
\langle V_y^2 \rangle&=&{\gamma_0^2 D\over \hat\tau_k c_0^2 F_0^2} \big[4-2(s^2+2s+2)e^{-s} \big], \label{Vy2c}
\\
\langle A_y^2 \rangle&=&{\gamma_0^2 D\over \hat\tau_k c_0^2 F_0^2} \big[1-{1\over2}(s^2-2s+2)e^{-s} \big] . \label{Ay2c}
\ea
In the long time limit and in the original variables, evolve as follows:
\be 
\langle X^2\rangle =2D\left(t - {11\over 8} \tau_r\right), \qquad 
\langle V^2\rangle={k_{_B}T\over m},\qquad  \langle A^2\rangle={1 \over
J\, \tau_r^2} \langle V^2\rangle_{eq}
\label{msdxva2} \ee
Here, the variance $\langle X^2\rangle$ is consistent with the one given in Eq. (\ref{msdxva}) if $J={1\over 4}$. In this case, the numerical simulation of Eq. (\ref{sale2}) are carried out taking into account the corresponding expression of each parameter $J$, $\hat\tau_k$, $y$, and $s$. 
See the details in Sec IV.
\begin{figure}[h!]
    \centering
    \includegraphics[scale=0.6]{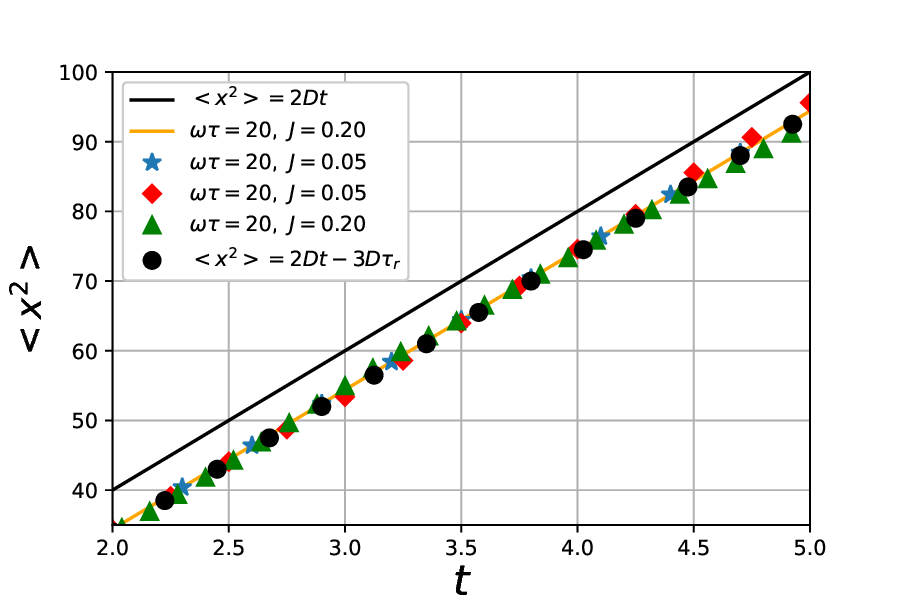}
    \caption{Mean Square Displacement (MSD), compared with both the Markovian one and numerical simulation.  Black line is the Markovian result, Orange line and blue stars the theoretical results  (\ref{Y2}) in the original variable $X$, for one value of $\omega\tau=20$ and two different values of $J=0.2, \, 0.05$, respectively. 
    Red diamonds and green triangles are  numerical simulation of Eq. (\ref{sale2}), for $\omega\tau=20$ and $J=0.05, \, 0.2$, respectively. Black dots represent the theoretical approximation (\ref{x22})    }
    \label{msd1}
\end{figure}

\begin{figure}[h!]
    \centering
    \includegraphics[scale=0.6]{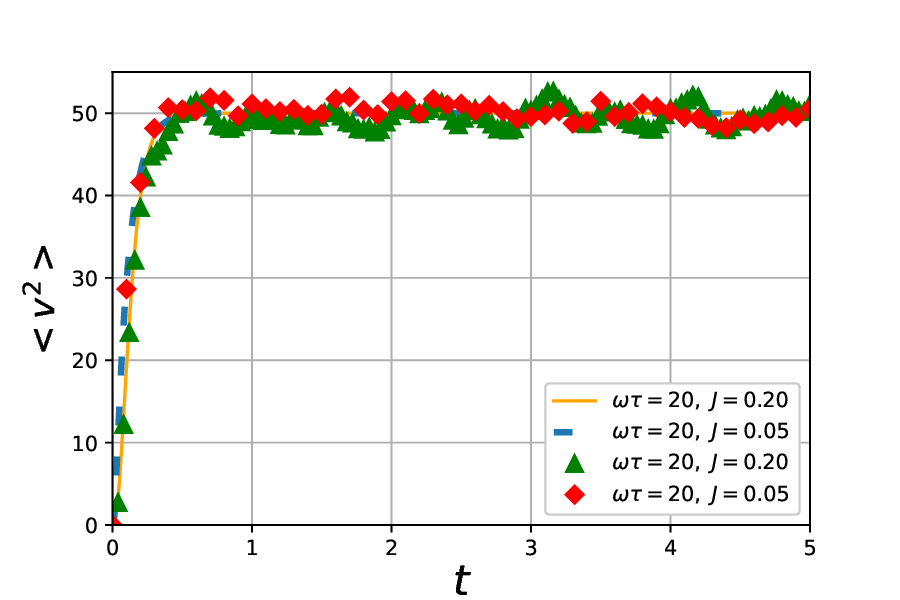}
    \caption{Velocity variance or Mean Square Velocity $\langle V^2\rangle$ compared with the numerical simulation. Orange line and blue dashed line are the theoretical results (\ref{Vy2}), in the original variable $V$, for a specific value of $\omega\tau=20$ and two values of $J=0.2, \, 0.05$, respectively. The numerical simulation is plotted for the same values of $\omega\tau$ and $J$. As the time gets large, all the curves tend toward the equilibrium value ${k_{_B}T\over m}$.}
    \label{msv1}
\end{figure}
\section{Comparison with the numerical simulation}

In this section we compare the explicit solutions reported in Sec. III with those derived from the numerical simulations of Eq. (\ref{sale2})
\begin{figure}[ht]
    \centering
    \includegraphics[scale=0.6]{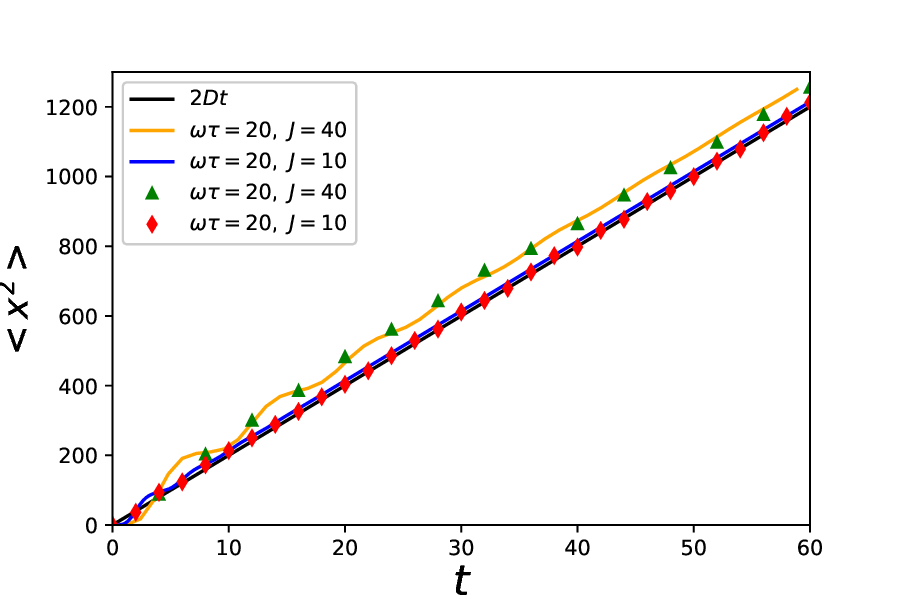}
    \caption{MSD $\langle X^2\rangle$, compared with both the Markovian result and  numerical simulation. Black line is the Markovian MSD. The orange and blue colors correspond to oscillatory behavior of Eq. (\ref{Y2a}) in the original variable $X$, for a specific value of $\omega\tau=20$ and two values of $J=40, \, 10,$ respectively. Both curves are attenuated taking the shape straight lines, and are above of the Markovian result. The green triangles and red diamonds are the numerical simulation, for the same values of $\omega\tau$ and $J$. }
    \label{msd2}
\end{figure}
\begin{figure}[h!]
    \centering
    \includegraphics[scale=0.6]{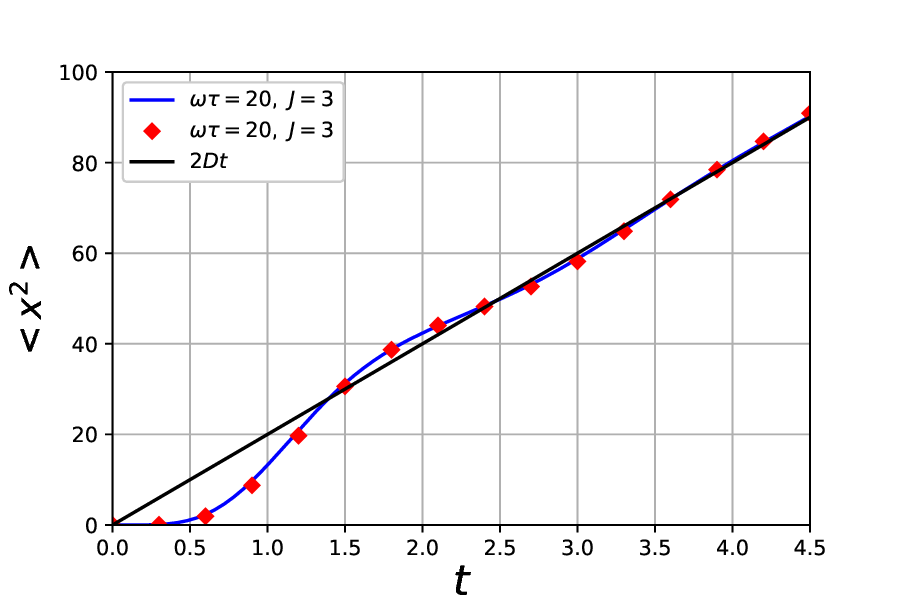}
    \caption{MSD $\langle X^2\rangle$, compared with both the Markovian result and  numerical simulation. Black line is the Markovian MSD. The blue color correspond to oscillatory behavior of Eq. (\ref{Y2a}) in the original variable $X$, for specific  values of $\omega\tau=20$ and $J=3$. It is shown that, in the long time limit, the oscillatory behavior is attenuated as a straight line towards the Makovian result. The red diamonds are the numerical simulation.}
    \label{msd3}
\end{figure}

$\bullet$ In Fig. \ref{msd1}, we show the comparison between the  variance or mean square displacement (MSD) $\langle X^2\rangle$ and both the theoretical result (\ref{x22}) and numerical simulation of Eq. (\ref{sale2}), for two specific values of $J$ in the interval $0<J<0.25$. As can be seen, in this interval the variance  (\ref{Y2}) for the original variable $X$ is practically the same as Eq. (\ref{x22}), and therefore the diffusion process is $J$-independent. In this case the variance (\ref{x22}) is the same as the Markovian one except for a time delay $t-{3\over 2}\tau_r$.

\begin{figure}[h]
    \centering
    \includegraphics[scale=0.6]{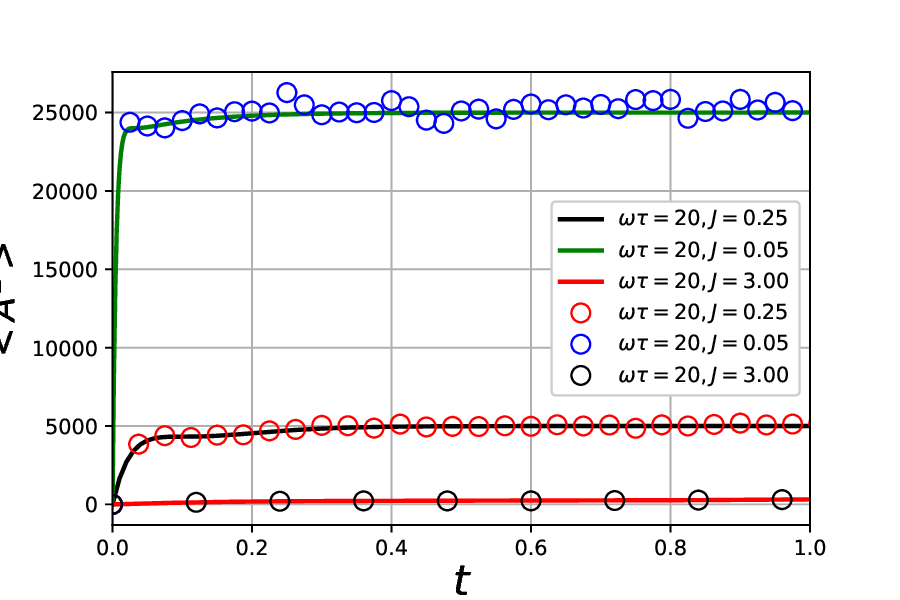}
    \caption{Mean square acceleration $\langle A^2\rangle$ for fixed value of $J=0.25$ and a value of $\omega\tau=20$.
    Green line corresponds to Eq. (\ref{Ay2}), black line to eq. (\ref{Ay2a}), and orange line to (\ref{Ay2c}), all in the original variable $A$. The circles are the corresponding simulation results.}
    \label{msa1}
\end{figure}

$\bullet$ The plot given in Fig. \ref{msv1} corresponds to the comparison between the theoretical velocity variance $\langle V^2\rangle$ and the numerical solution of Eq. (\ref{sale2}).  
Both results show that the velocity reaches its equilibrium expected result $\langle V^2\rangle_{eq}={k_{_B}T\over m}$, as expected. The variance $\langle A^2\rangle$ in the original variable has a similar dynamic behavior and  plotted in Fig. \ref{msa1}, for $J=0.05$ and compared with the numerical simulation of Eq. (\ref{sale2}).  Its equilibrium value is consistent with the result $\langle V^2\rangle_{eq}/J \tau_r^2$.

$\bullet$ In Fig. \ref{msd2},  we plot the theoretical variance $\langle X^2\rangle$ coming from Eq. (\ref{Y2a}), for two values of $J=10, 20$ ( recall that $J>0.25$). In this case the oscillatory behavior is due to complex nature of the roots $\lambda_2$ and $\lambda_3$. As can be seen, as time is longer, the oscillatory behavior becomes attenuated taking the shape of a straight line. This can be corroborated because in the long time limit the variance $\langle X^2\rangle$ is the straight line given in Eq. (\ref{msdxva}). In this limit case, a remarkable result is obtained when $J=3$, because $\langle X^2\rangle=2Dt$, which is the Markovian result. This fact is clearly seen in Fig. \ref{msd3}, in which the oscillatory variance follows the Markovian straight line, as well as the numerical simulation results.

\begin{figure}[h]
    \centering
    \includegraphics[scale=0.6]{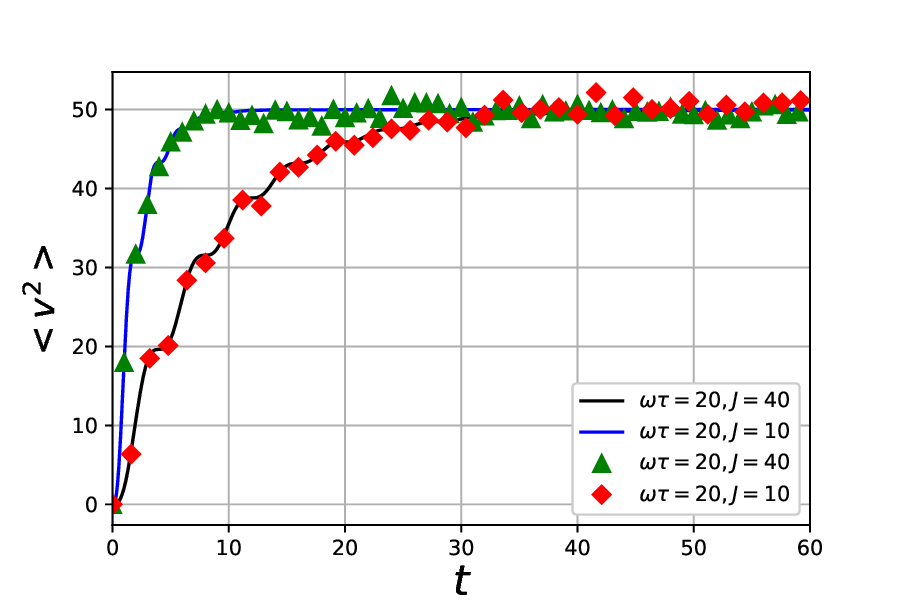}    \caption{Velocity variance $\langle V^2\rangle$ compared with the numerical simulation. Blue and black lines are the theoretical results (\ref{Vy2a}), in the original variable $V$, for a specific value of $\omega\tau=20$ and two values of $J=40, \, 10$, respectively. The numerical simulation is plotted for the same values of $\omega\tau$ and $J$.   In the long time limit, all the curves tend toward the equilibrium value ${k_{_B}T\over m}$, as expected. }
    \label{msv2}
\end{figure}

$\bullet$ In the case of complex roots, the theoretical velocity variance $\langle V^2\rangle$ coming from Eq. (\ref{Vy2a}), is plotted in Fig. \ref{msv2}, where it is compared with the numerical simulation. Also its oscillatory behavior at early times, as well as its equilibrium value ${k_{_B}T\over m}$ as the time goes out, are shown.  The variance $\langle A^2\rangle$ is plotted in Fig. \ref{msa1} for the value of $J=3.0$ and also compared with the numerical simulation. With values close to zero.  

\begin{figure}[h]
    \centering
    \includegraphics[scale=0.6]{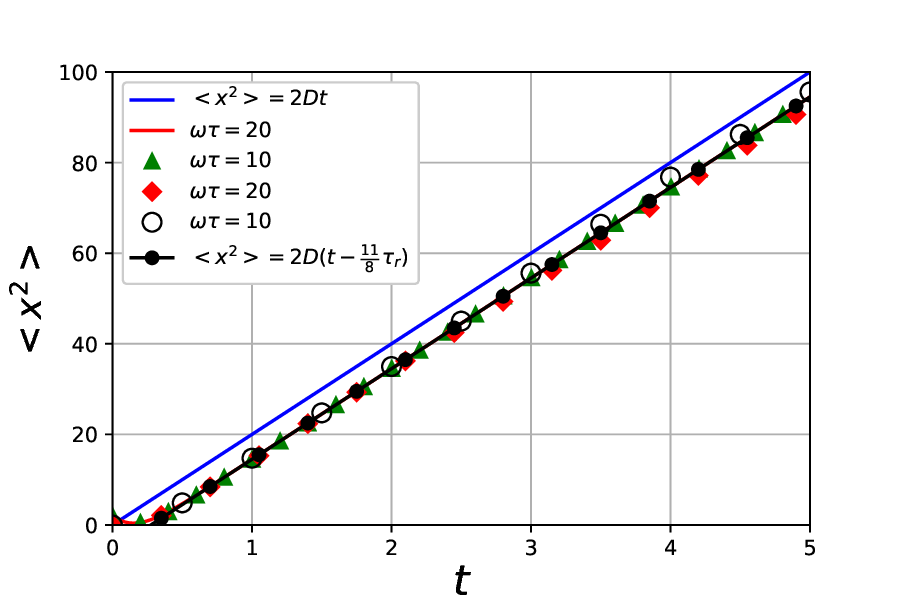}
    \caption{MSD $\langle X^2\rangle$ 
    compared with both the Markovian result and  numerical simulation. The blue straight line is the Markovian MSD. The red line and green triangles are the theoretical variance given by Eq. (\ref{Y2c}) in the original variable $X$, for fixed value of $J=0.25$, and two values of $\omega\tau=20, \, 10$. Red diamonds and circles correspond to the numerical simulation, and black dots the MSD given in Eq. (\ref{msdxva2}) }
    \label{msd4}. 
\end{figure}

$\bullet$ In the critical case $J={1\over 4}$ and thus $\hat\tau_k={1\over 4}\tau_r$, the theoretical variance $\langle X^2\rangle$ coming from Eq. (\ref{Y2c}), is plotted in Fig. \ref{msd4}. In this case, it is shown that no matter what the value of $\omega\tau$ is, both the theoretical and numerical simulation results coincide. Moreover, they also coincide with the Markovian result except for a time delay $t-{11\over 8}\tau_r$. 

\begin{figure}[h]
    \centering
    \includegraphics[scale=0.6]{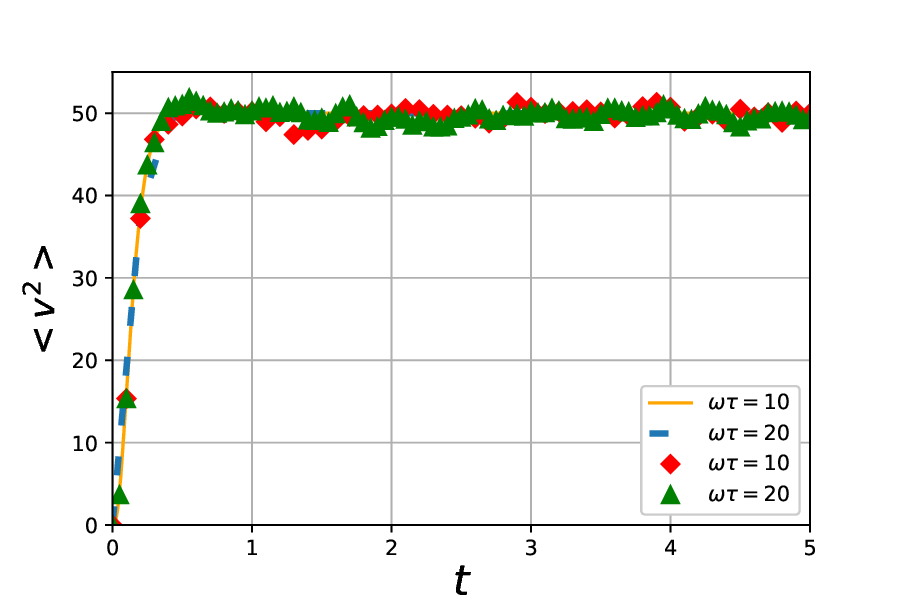}
    \caption{Mean square velocity $\langle V^2\rangle$, given by Eq. (\ref{Vy2c}) in the original variable $X$, for fixed value of $J=0.25$ and two values of $\omega\tau=10, \, 20$. The orange and blue lines are the theoretical results; red diamonds and  green triangles are the numerical simulations. All the results coincide no matter what the value of $\omega\tau$ is. }
    \label{msv3}
\end{figure}

$\bullet$ In Fig. \ref{msv3}, 
the variance $\langle V^2\rangle$ coming from Eq. (\ref{Vy2c}) is compared with the numerical simulation results. Both results are consistent and tend towards 
the equilibrium value ${k_{_B}T\over m}$ as it should be. The variance $\langle A^2\rangle$ is also plotted in Fig. \ref{msa1} for the value of $J=0.25$ and compared with the numerical simulation. It is clearly shown that, as $J$ is greater the value of acceleration variance is lesser.


\newpage
\section{Concluding Remarks}

In this work we have been able to describe the statistic of the electronic plasma diffusion in a classic way,  taking into account the radiation reaction force. This has successfully been achieved by means of a GLE under the action of a time-dependent electric field, in the form of an exponential decay. The proposed electric field induces an electric force on the electron,  which is capable of producing a delay in the electron's characteristic time. The theoretical approach allows to establish a quasi-Markovian third-order time derivative stochastic differential equation, containing a friction force $m(\tau-k\tau_e)\dddot x$ in which the effective memory time $\hat\tau_k=\tau-k\tau_e$ is required  to be positive and finite in order to avoid runaway solutions.

The electron gas statistic is better calculated using the dimensionless stochastic differential equation, in which the $J$ parameter plays an important role. For the real and critical roots, the effects of the radiation reaction force through the $J$ parameter, appear only in the acceleration variance, as shown in Fig. \ref{msa1}. However, in the case of complex roots, ($J>{1\over 4}$, \, or \, $\hat\tau_k>{1\over 4}\tau_r$),  the influence of the radiation reaction force is more noticeable in the variance $\langle X^2\rangle$ than the acceleration one. In this case, the MSD exhibits an oscillatory behavior whose amplitude of oscillation decreases as time progresses, taking the shape of a straight line. The influence of the radiation reaction force still remains in the long time limit for which the $\langle X^2\rangle$ is indeed a straight line, given by $\langle X^2\rangle=2D(t- {J-3\over 2}\tau_r)$, (see Figs. \ref{msd2} and \ref{msd3} ). In this limiting case, the mean square displacement MSD  is the same as the Markovian one except for a time delay $t-{J-3\over 2}\tau_r$. In the case of real roots, the variance $\langle X^2\rangle$ does not depend on $J$ and it is given by Eq. (\ref{x22}). In the critical case, for a given value of $J=0.25$, the mean square displacement is also independent of the value of $\omega\tau$ and  all coincide with one given in Eq. (\ref{msdxva2}). Lastly, according to the results reported in Fig. \ref{msa1}, the acceleration variance  decreases as $J$ increases, and reaches its equilibrium value faster than the velocity variance. \\         

\noindent{\bf Acknowledgments}

J.F.G.C. thanks to CONAHCyT for a grant in a Postdoc position at Metropolitan Autonomus University, campus Iztapalapa. 
Also, O.C.V thanks to CONAHCyT for a grant. GAP and NSS thank COFAA, EDI IPN and CONAHCyT.

\bibliographystyle{unsrtnat}
\bibliography{biblio}

\end{document}